\documentclass[12pt,headings=big,numbers=noenddot,DIV=14,a4paper]{scrartcl}%
\pdfoutput=1

\usepackage{amsmath,amssymb,amsfonts}
\usepackage[normal,font=small,labelfont=bf,labelsep=period]{caption}
\usepackage[pdftex]{color,graphicx}
\usepackage[english]{babel}
\usepackage[compress]{cite}
\addtokomafont{disposition}{\rmfamily\boldmath}
\usepackage{graphicx}
\usepackage{subfig}
\usepackage{arydshln}
\usepackage{slashed} 

\usepackage[dvipsnames]{xcolor}
\definecolor{darkblue}{rgb}{0,0.2,0.6}
\definecolor{darkgreen}{rgb}{0,0.4,0}

\usepackage[linktoc=page,bookmarks=false,colorlinks=false,
linkbordercolor=RoyalBlue,citebordercolor=ForestGreen,urlbordercolor=CornflowerBlue]{hyperref}

\numberwithin{equation}{section}
\newcommand{\arXhref}[1]{\href{http://arxiv.org/abs/#1}{#1}}

\newcommand{\be} {\begin{equation}}
\newcommand{\ee} {\end{equation}}
\newcommand{\ba} {\begin{eqnarray}}
\newcommand{\ea} {\end{eqnarray}}

\newcommand{\cB} {\mathcal B}

\newcommand{\GeV}{\text{GeV}}

\DeclareFontFamily{OT1}{pzc}{}
\DeclareFontShape{OT1}{pzc}{m}{it}{<-> s * [1.10] pzcmi7t}{}
\DeclareMathAlphabet{\mathpzc}{OT1}{pzc}{m}{it}

\begin{document}

\begin{flushright}
 ZU-TH-44/15 \\
\end{flushright}

\thispagestyle{empty}

\begin{center}
\vspace{1.5cm}
    {\Large\bf  Anomalies in $B$-decays and $U(2)$ flavour symmetry} \\[1cm]
   {\bf Riccardo Barbieri$^{a,d}$, Gino Isidori$^{b,c}$, 
 Andrea Pattori$^{b,e}$, Fabrizio Senia$^d$ }   \\[0.5cm]
    {\em $(a)$ Institute of Theoretical Studies, ETH Z\"urich, CH-8092   Z\"urich, Switzerland}\\
  {\em $(b)$  Physik-Institut, Universit\"at Z\"urich, CH-8057 Z\"urich, Switzerland}  \\ 
  {\em $(c)$  INFN, Laboratori Nazionali di Frascati, I-00044 Frascati, Italy}  \\
   {\em $(d)$  Scuola Normale Superiore and INFN, Piazza dei Cavalieri 7, 56126 Pisa, Italy}\\
   {\em $(e)$  Dipartimento di Fisica e Astronomia `G. Galilei', Universit\`a di Padova, Via Marzolo 8, I-35131 Padua, Italy}
\end{center}

\centerline{\large\bf Abstract}
\begin{quote}
\indent
The collection of a few anomalies in semileptonic $B$-decays invites to speculate about the emergence of some strikingly new phenomena. Here we offer a possible interpretation of these anomalies in the context of a weakly broken $U(2)^5$ flavour symmetry
and lepto-quark mediators.
\end{quote}
\vspace{5mm}

\newpage

\section{Introduction}

In the last years quite a few anomalies in semileptonic B-decays have emerged. While individually any of them requires confirmation before being taken under serious consideration, their collection motivates to speculate about the possible emergence of some strikingly new phenomena. The purpose of this paper is to offer a possible interpretation of these anomalies in the context of a weakly broken $U(2)^n$-flavour symmetry. 

The observations we refer to have received and are receiving a lot of attention. 
The deviations from the SM that are both statistically more significant and whose theoretical error is 
negligible (compared to the present experimental error) can be summarized as follows:

\begin{itemize}
\item 
An overall  $3.9 \sigma$ deviation from $\tau/l$ universality $(l = \mu,e)$ in charged current semileptonic $B\rightarrow D^{(*)}$ decays~\cite{Lees:2013uzd,
Huschle:2015rga,Aaij:2015yra}:\footnote{~The results in eq.~(\ref{R_D})
are obtained using the theory predictions $\cB(B \to D^* \tau \nu)/\cB(B \to D^* \ell \nu )_{\rm SM} = 0.252\pm 0.003$~\cite{Fajfer:2012vx}
and  $\cB(B \to D  \tau \nu)/\cB(B \to D \ell \nu )_{\rm SM} =  0.300\pm 0.008$~\cite{Na:2015kha}. }
\begin{equation}
R^{\tau/l}_{D^{(*)}}=\frac{\mathcal{B}(\bar{B}\rightarrow D^{(*)}\tau\bar{\nu})/\mathcal{B}(\bar{B}\rightarrow D^{(*)}\tau\bar{\nu})_{SM}
}{\mathcal{B}(\bar{B}\rightarrow D^{(*)} l\bar{\nu})/\mathcal{B}(\bar{B}\rightarrow D^{(*)} l\bar{\nu})_{SM}},
\end{equation}
\begin{equation}
R_D^{\tau/l}= 1.37\pm 0.17,  \quad \quad
R_{D^*}^{\tau/l} =  1.28\pm 0.08
\label{R_D}
\end{equation}

\item 
 A $2.6 \sigma$ deviation from $\mu/e$ universality in the neutral current $b\rightarrow s$ transition~\cite{Aaij:2014ora}: 
\begin{equation}
R_K^{\mu/e} = \frac{\mathcal{B}(B\rightarrow K\mu^+\mu^-)}{\mathcal{B}(B\rightarrow Ke^+e^-)} =0.745^{+0.090}_{-0.074}\pm0.036
\label{R_K}
\end{equation}
predicted to be one in the Standard Model (SM) with better than $1\%$ accuracy.
\end{itemize}
This last neutral current anomaly may be related to other tensions with the SM in the branching ratios and in the 
angular distributions of semileptonic $b\rightarrow s$ transitions, particularly in $B\rightarrow K^{(*)}\mu^+\mu^-$ and $B\rightarrow \phi\mu^+\mu^-$ 
(see Ref.~\cite{Descotes-Genon:2015uva,Altmannshofer:2014rta} for an updated discussion).

The interpretation of a $30 \%$ deviation from the SM in a tree level charged current interaction, eq. (\ref{R_D}), calls for an exchange capable to produce at low energy an effective 4-fermion interaction proportional to the operator $(\bar{c}_L\gamma_\mu b_L)(\bar{\tau}_L\gamma_\mu \nu_L)$. On these grounds one may want to interpret the neutral current anomaly in eq. (\ref{R_K})      as also  due  to an  LL operator of the form $(\bar{s}_L\gamma_\mu b_L)(\bar{\mu}_L\gamma_\mu \mu_L)$. Although not exclusively, it is known that such an operator can as well improve the fit in the angular distribution of the semileptonic $b\rightarrow s$ transitions~\cite{Descotes-Genon:2015uva,Altmannshofer:2014rta}. There is however a problem to face. While both anomalies hint at a $20\div 30\%$ deviation from the SM, there is an important difference among them. The charged current anomaly is a deviation from a SM tree level amplitude involving the third generation of leptons, whereas the neutral current one is a putative correction to a SM loop effect only concerning the first two generations of leptons. 

This motivates us to ask whether there is a flavour group $\mathcal{G}_F$ and a tree level exchange $\Phi$ such that:
\begin{itemize}
\item 
With unbroken $\mathcal{G}_F$, $\Phi$ couples to the third generation of quarks and leptons only;
\item
After $\mathcal{G}_F$-breaking, the needed operators, as mentioned in the previous paragraph, are generated as a small perturbation.

\end{itemize}
The answer is positive with 
\begin{equation}
\mathcal{G}_F = \mathcal{G}_F^q\times \mathcal{G}_F^l
\end{equation}
\begin{equation}
\mathcal{G}_F^q = U(2)_{Q}\times U(2)_{u} \times U(2)_{d}\times U(1)_{d3}, \quad\quad
\mathcal{G}_F^l= U(2)_{L}\times U(2)_{e} \times U(1)_{e3}~,
 \end{equation}
 in the notation of Ref.~\cite{Barbieri:2011ci,MFV}, and $\Phi$ is a leptoquark singlet under  
 $\mathcal{G}_F$, carrying either one of the following quantum numbers under the SM gauge group:
 \begin{itemize}
 \item[1.]  $U_\mu = (3,1)_{2/3}$, Vector singlet-model; 
 \item[2.] $\mathcal{U}_\mu = (3,3)_{2/3}$, Vector triplet-model;
 \item[3.] $S = (\bar{3}, 3)_{1/3}$, Scalar triplet-model.
 \end{itemize}

Dynamical explanations of the above anomalies have already been proposed in the literature
both in terms of vector uncoloured mediators~\cite{Greljo:2015mma}  and in terms of leptoquark mediators~\cite{Calibbi:2015kma,Fajfer:2015ycq,Bauer:2015knc}.\footnote{~Leptoquark explanations 
of a single set of anomalies (either neutral or charged currents) have been discussed in Ref.~\cite{LQ}.
For a recent discussion of the two set of anomalies in terms of effective four-fermion operators see Ref.~\cite{EFT}.}
Here we focus on the specific realization  of leptoquark models  based on the flavor group $\mathcal{G}_F$ since: 

\begin{itemize}
\item[i)] this  explains their dominant coupling to the third generation only (in particular only to the left-handed quark and lepton doublets which are the only 
$\mathcal{G}_F$-invariant fermions);
\item[ii.)]  the breaking of $\mathcal{G}_F$  specifies the source of the flavour violating couplings needed to give rise predominantly to the operator $(\bar{c}_L\gamma_\mu b_L)(\bar{\tau}_L\gamma_\mu \nu_L)$ and, at a weaker level, to $(\bar{s}_L\gamma_\mu b_L)(\bar{\mu}_L\gamma_\mu \mu_L)$ as well.
\end{itemize}
About the needed breakings of $\mathcal{G}_F$, we stick to the  minimal set of spurions
\begin{equation}
y_{d3}=(1,1,1)_{-1}\quad\quad \Delta_{u}=(2,\bar{2},1)_{0} \quad\quad \Delta_{d}=(2,1,\bar{2})_{0}\quad\quad \mathbf{V}_Q=(2,1,1)_{0}
\label{spur3}
\end{equation}
for $\mathcal{G}_F^q$ \cite{Barbieri:2011ci} and
\begin{equation}
y_{e3}=(1,1)_{-1}\quad\quad \Delta_{e}=(2,\bar{2})_{0} \quad\quad \mathbf{V}_L=(2,1)_{0}
\label{spur2}
\end{equation}
for $\mathcal{G}_F^l$\cite{Barbieri:2012uh,Blankenburg:2012nx}.

In the following we write down in the three cases above the leptoquark (LQ) couplings to the fermions in their physical bases after inclusion of $\mathcal{G}_F$-breaking (Section 2) and we calculate the relevant amplitudes  at tree level (Section 3). Consistency with current data is achieved only by the $U_\mu = (3,1)_{2/3}$ model above.
The dominant loop effects when the tree level amplitude vanishes are calculated in Section 4 for the surviving vector-singlet model. The overall consistency of the model with data  is illustrated in Section 5, where further expected signals are also examined. A tentative UV completion of the phenomenological model is briefly outlined in Section 6. Summary and conclusions are drown in Section 7.

\section{Leptoquark Lagrangians after $\mathcal{G}_F$-symmetry breaking}

The couplings of the three LQ to the SM electroweak doublets $Q_{L}$ and $L_{L}$  are ($a$ is an index in the vector representation of $SU(2)_L$, whereas the colour indices are left understood)
\begin{equation}\begin{split}
\mathcal{L}_1=g_{U}(\bar{Q}_{L}\gamma^{\mu}F L_{L})U_{\mu}+\text{h.c}
\label{L1}
\end{split}\end{equation}
\begin{equation}\begin{split}
\mathcal{L}_2=g_{\vec{U}}(\bar{Q}_{L}\gamma^{\mu}\frac{\sigma^a}{2}F L_{L})U^a_{\mu}+\text{h.c}
\end{split}\end{equation}
\begin{equation}\begin{split}
\mathcal{L}_3=g_{\vec{S}}(\bar{Q}^c_{L}\frac{\sigma^a}{2}F (i\sigma^2L_{L}))S^a+\text{h.c};\quad\quad
Q_{L}^{c}\equiv(Q_{L})^{c}
\end{split}\end{equation}
where $F$ is a matrix in flavour space which, in the $\mathcal{G}_F$ symmetric limit, is $F_{ij} = \delta_{i3}\delta_{j3}$.
On the other hand, after symmetry breaking along the directions (\ref{spur3}) and  (\ref{spur2}), $F$ takes the form
\begin{equation}
F_{ij}=\delta_{i3}\delta_{j3}+a V_{Q_i}\delta_{j3}+b\delta_{i3}V_{L_j}+c V_{Q_i}V_{L_{j}}
\label{Fij}
\end{equation}
where, by symmetry transformations, we can write
\begin{equation}
V_{Q}=\begin{pmatrix}
          0\\
          \epsilon_Q\\
          0
         \end{pmatrix}\quad\quad\quad
V_{L}=\begin{pmatrix}
          0\\
          \epsilon_l\\
          0
         \end{pmatrix}     
\end{equation}         
in terms of two small real parameters  $\epsilon_Q, \epsilon_L$ and $a,b,c$ are arbitrary coefficients, generically of order unity, that we shall also take real for simplicity.      In eq. (\ref{Fij}) we are neglecting terms proportional    
to the product $y_b y_\tau$ of the bottom and $\tau$ Yukawa couplings or to products of $\Delta_{u,d,e}$. At the same time and in the same bases as eq. (\ref{Fij}), the Yukawa matrices for the charged fermions take the form
\begin{equation*}
Y_{u}=\left(
\begin{array}{c|c}
\Delta_{u} & y_{t}\mathbf{V_Q} \\ \hline
0 & y_{t}.
\end{array}\right)
\quad\quad
Y_{d}=\left(
\begin{array}{c|c}
\Delta_{d} & y_{b}x_{b}\mathbf{V_Q} \\ \hline
0 & y_{b}
\end{array}\right)
\quad\quad
Y_{e}=\left(
\begin{array}{c|c}
\Delta_{e} & y_{\tau}\mathbf{V_L} \\ \hline
0 & y_{\tau}
\end{array}\right)
\end{equation*}
where $x_b$ is another unknown coefficient in general of order unity.

One goes to the physical bases for all the charged fermions by an approximate diagonalization of these Yukawa matrices. In these physical bases the 
LQ interaction Lagrangians acquire the form
\begin{equation}\label{e1}
\mathcal{L}_1=g_{U}
(\bar{u}_L\gamma^{\mu}F^{U}\nu_L+\bar{d}_{L}\gamma^{\mu}F^{D}e_{L})U_{\mu}+\text{h.c}
\end{equation}
\begin{equation}
\mathcal{L}_2=\frac{g_{\vec{U}}}{\sqrt{2}}\Big[\frac{1}{\sqrt{2}}
(\bar{u}_L\gamma^{\mu}F^{U}\nu_L-\bar{d}_{L}\gamma^{\mu}F^{D}e_{L})U^{2/3}_{\mu}+
(\bar{u}_L\gamma^{\mu}F^{U}e_{L})U_{\mu}^{5/3}+(\bar{d}_L\gamma^{\mu}F^{D}\nu_L)U_{\mu}^{-1/3}\Big]+\text{h.c}
\end{equation}
\begin{equation}
\mathcal{L}_3=
\frac{g_{\vec{S}}}{\sqrt{2}}\Big[\frac{1}{\sqrt{2}}
(\bar{u}^c_LF^{U} e_L+\bar{d}^c_{L} F^{D}\nu_L)S^{1/3}+
(\bar{u}^c_LF^{U}\nu_L)S^{-2/3}+(\bar{d}^c_L F^{D} e_L)S^{4/3}\Big]+\text{h.c}
\end{equation}
where
\begin{equation}
F^U  =\begin{pmatrix}
V_{ub}(s_l\epsilon_l)A_u &   V_{ub}(c_l\epsilon_l)A_u& V_{ub}(1-a)r_u\\     
V_{cb}(s_l\epsilon_l)A_u &   V_{cb}(c_l\epsilon_l)A_u& V_{cb}(1-a)r_u\\ 
V_{tb}(s_l\epsilon_l)(b-1) &   V_{tb}(c_l\epsilon_l)(b-1)& V_{tb}\\ 
\end{pmatrix}
\end{equation}
\begin{equation}
F^D =\begin{pmatrix}
V_{td}(s_l\epsilon_l)A_d &   V_{td}(c_l\epsilon_l)A_d& V_{td}[1-(1-a)r_u]\\     
V_{ts}(s_l\epsilon_l)A_d &   V_{ts}(c_l\epsilon_l)A_d& V_{ts}[1-(1-a)r_u]\\ 
V_{tb}(s_l\epsilon_l)(b-1) &  V_{tb}(c_l\epsilon_l)(b-1)& V_{tb}\\ 
\end{pmatrix}
\end{equation}
\begin{equation}
  r_u=\frac{1}{1-x_b}\quad\quad A_u=r_u(b-1+a-c) \quad\quad A_d=b-1-A_u 
\end{equation}
and  $\theta_l$ is the angle $(s_l =\sin{\theta_l}, c_l = \cos{\theta_l})$ in the unitary transformation that diagonalizes $\Delta_l$ on the left side.
We are working in the basis of neutrino current-eigenstates, where the charged current leptonic weak interactions are flavour-diagonal. We are also neglecting a phase in the $1-3, 3-1$ elements of these flavour matrices since it does not play any role in our following considerations.

\section{Tree-level amplitudes}

In the three models defined in Section 1 the tree-level exchanges of the corresponding LQ give rise to effective Lagrangians relevant to 
charged-current and neutral-current semileptonic $B$ and $K$ decays. For $b\rightarrow c  \tau  \bar\nu_3$ one has  
\begin{equation}
\mathcal{L}_{eff}^{b\rightarrow c\tau\nu}= (-\frac{g_U^2}{M_U^2},  \frac{g_{\vec{U}}^2}{4M_{\vec{U}}^2}, \frac{g_{\vec{S}}^2}{8M_{\vec{S}}^2}) r_u V_{cb} (1-a) (\bar{c}_L\gamma_\mu b_L)(\bar{\tau}_L\gamma_\mu \nu_{3L})
\label{Leff_cc}
\end{equation}
to be compared with the SM result
\begin{equation}
\mathcal{L}_{SM}^{b\rightarrow c\tau\nu}= -\frac{g^2}{2M_W^2}  V_{cb} (\bar{c}_L\gamma_\mu b_L)(\bar{\tau}_L\gamma_\mu \nu_{3L}).
\label{Leff_cc_SM}
\end{equation}
In both cases (SM and LQ exchange) the $b \rightarrow u \tau  \bar\nu_3$ effective Lagrangians are obtained from the above ones with the 
exchange $c\to u$ and $V_{cb} \to V_{ub}$. 

For the neutral-current processes $b\rightarrow s  \ell \bar\ell$, with $\ell=e,\mu,\tau$,  the LQ exchange gives
\ba
\mathcal{L}_{eff}^{b\rightarrow s\mu\mu} &=&  (-\frac{g_U^2}{M_U^2}, - \frac{g_{\vec{U}}^2}{4M_{\vec{U}}^2}, \frac{g_{\vec{S}}^2}{4M_{\vec{S}}^2}) V_{tb} V^*_{ts} (\bar{s}_L\gamma_\mu b_L) \Big[ (1-(1-a)r_u)(\bar{\tau}_L\gamma_\mu \tau_{L}) \nonumber \\
&&  + (c_l\epsilon_l)^2 (b-1)A_d (\bar{\mu}_L\gamma_\mu \mu_{L})  +  (s_l\epsilon_l)^2 (b-1)A_d (\bar{e}_L\gamma_\mu e_{L})    \Big]~,
\label{eq:FCNCbsll}
\ea
whereas the lepton-universal local $b\rightarrow s \ell \bar\ell$ effective interaction present in the SM reads 
\begin{equation}
\mathcal{L}_{SM}^{b\rightarrow s \ell \ell} \approx -\frac{8 G_F}{\sqrt{2}}V_{tb}V_{ts}^*\frac{\alpha}{4 \pi} C_9^{SM}(\bar{s}_L\gamma_\mu b_L)(\bar{\ell}_L\gamma_\mu \ell_{L})~, 
\end{equation}
with $C_9^{SM} \approx 4.2$.
Finally, for  $b\rightarrow s \nu_3 \bar \nu_3$  and $s\rightarrow d \nu_3 \bar \nu_3$   amplitudes,  the   LQ exchange  gives
\ba
\mathcal{L}_{eff}^{b(s)\rightarrow s(d)\nu\nu} &=& ( 0 ,  - \frac{g_{\vec{U}}^2}{2M_{\vec{U}}^2}, \frac{g_{\vec{S}}^2}{8M_{\vec{S}}^2}) 
(\bar{\nu}_{3L}\gamma_\mu \nu_{3L})  \Big[ V_{tb} V^*_{ts} ((1 +r_u (a-1))  (\bar{s}_L\gamma_\mu b_L) \nonumber \\
&& + V_{ts} V^*_{td} |1 +r_u (a-1)|^2  (\bar{d}_L\gamma_\mu s_L) \Big]~,
\label{Leff_nc}
\ea
to be compared with 
\begin{equation}
\mathcal{L}_{SM}^{b(s)\rightarrow s(d) \nu\nu} = -\frac{8 G_F}{\sqrt{2}} C_\nu^{SM} \frac{\alpha}{4 \pi} \left[ V_{tb}V_{ts}^* (\bar{s}_L\gamma_\mu b_L)
+ V_{ts}V_{td}^* (\bar{d}_L\gamma_\mu s_L) \right] \sum_{i=1}^3(\bar{\nu}_{iL}\gamma_\mu \nu_{iL})~,
\label{Leff_nc_SM}
\end{equation}
where $C_{\nu}^{SM} \approx -6.3$ (and we have omitted the sub-leading charm contribution in the  $s\rightarrow d \nu \bar{\nu}$ case).
Note the absence of a tree level contribution to $b\rightarrow s \nu \bar \nu$ from the $SU(2)$-singlet vector LQ (model 1), 
as noted first in Ref.~\cite{Calibbi:2015kma}.

\subsection{Tree-level constraints on the parameter spaces}

From eq.s (\ref{Leff_cc},\ref{Leff_cc_SM}), neglecting small corrections to $\mathcal{B}(\bar{B}\rightarrow D^{(*)} l\bar{\nu})$ suppressed by the $\epsilon_l$ factor, one has for the three models of Section 1
\begin{equation}
R^{\tau/l}_{D^{(*)}} \approx 
1+ (R_U, -\frac{1}{4} R_{\vec{U}}, -\frac{1}{8} R_{\vec{S}})r_u (1-a)
\end{equation}
where
\begin{equation}
(R_U, R_{\vec{U}},  R_{\vec{S}}) = \frac{4 M_W^2}{g^2} ( \frac{g_U^2}{M_U^2},  \frac{g_{\vec{U}}^2}{M_{\vec{U}}^2},  \frac{g_{\vec{S}}^2}{M_{\vec{S}}^2}).
\end{equation}
Similarly from eq.s (\ref{Leff_nc}, \ref{Leff_nc_SM}) one has
\begin{equation}
R_{K^{(*)}\nu} = \frac{\mathcal{B}(\bar{B}\rightarrow K^{(*)}\nu\bar{\nu})}{\mathcal{B}(\bar{B}\rightarrow K^{(*)}\nu\bar{\nu})_{SM}} \approx \frac{1}{3}\Big(3+2\text{Re}(x)+|x|^2\Big)             
\end{equation}
where for the various models it is
\begin{equation}
 (x_U, x_{\vec{U}},  x_{\vec{S}}) =-\frac{\pi}{\alpha c_{\nu}^{SM}}[ 1- r_u(1-a)]\Big( 0, -\frac{R_{\vec{U}}}{2},\frac{R_{\vec{S}}}{8}\Big).
\end{equation}
A similar formula holds also for $R_{\pi\nu\bar\nu}= \mathcal{B}(K\rightarrow \pi \nu\bar{\nu})/\mathcal{B}(K\rightarrow \pi \nu\bar{\nu})_{SM}$, 
in the limit where we neglect the subleading charm contribution and replace $[ 1- r_u(1-a)]$ with $[ 1- r_u(1-a)]^2$.

\begin{figure}[t]
\centering
\includegraphics[scale=0.85]{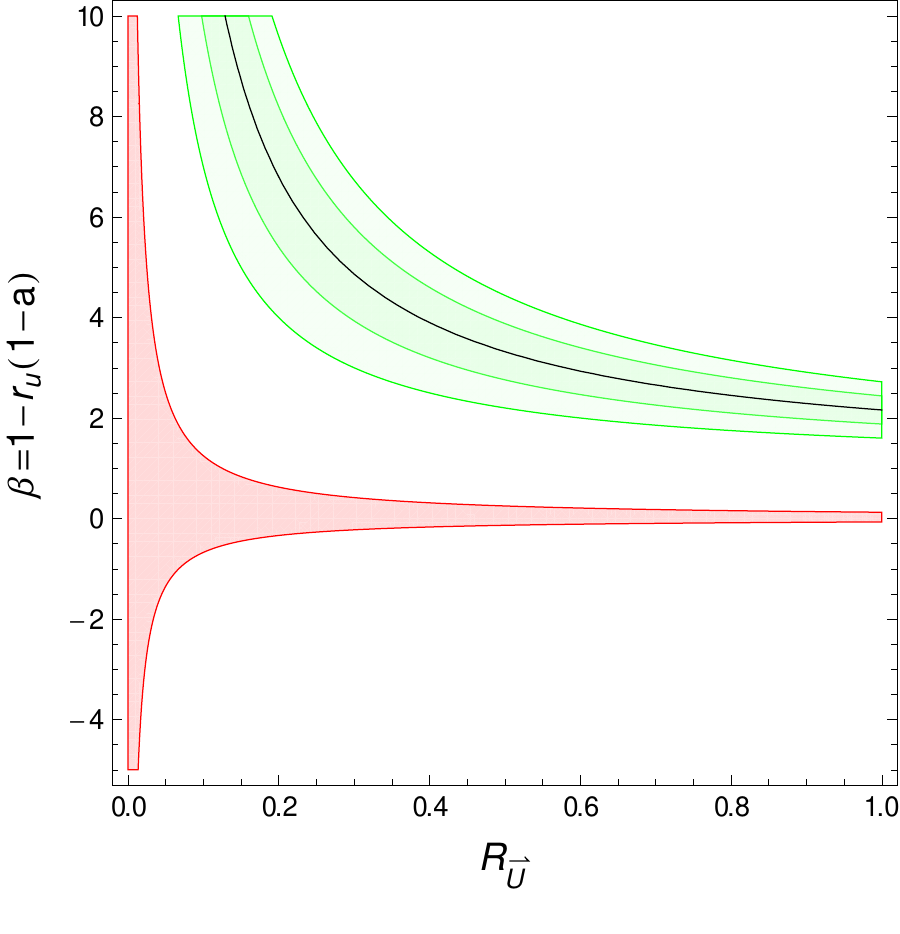}\hspace{7pt} 
\includegraphics[scale=0.85]{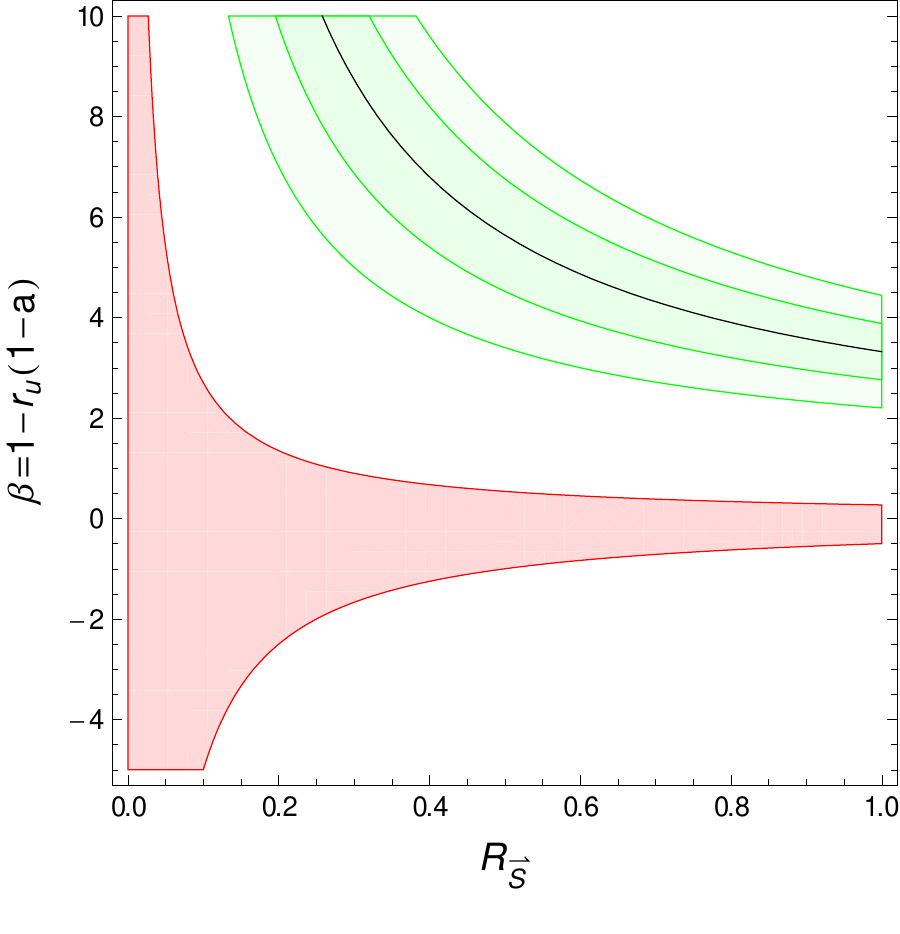} 
\caption{Allowed parameter spaces for the Vector-triplet $\vec{U}$ (left, $R_{\vec{U}} =4g_{\vec{U}}^2 M_W^2/g^2M_{\vec{U}}^2$) and for the Scalar-triplet $\vec{S}$ (right, $R_{\vec{S}} =4g_{\vec{S}}^2 M_W^2/g^2M_{\vec{S}}^2$
) from $R^{\tau/l}_{D^{(*)}}$ (at $1\sigma$ ($2\sigma$) darker (lighter) green region) and $R_{K^{(*)}\nu}$ (red region).}
\label{fig:tree}
\end{figure}

Unlike the case for $\mathcal{L}_{eff}^{b\rightarrow s\mu\mu}$, both $R^{\tau/l}_{D^{(*)}}$ and $R_{K^{(*)}\nu}$
depend on a single combination of flavour parameters $\beta = 1- r_u(1-a)$. 
This puts a strong constraint on models 2 and 3 of Section 1, shown in Fig. \ref{fig:tree}
(for $R_{K^{(*)}\nu}$ we use the bound  $R_{K^{(*)}\nu} < 4.3$~\cite{PDG}), making them highly disfavoured~\cite{Calibbi:2015kma}.
From now on we therefore concentrate our attention on model 1 with a Vector-singlet LQ. 

For later use, $R_{K}^{\mu/e}$ in this model is
\begin{equation}
 R_{K}^{\mu/e}=\frac{\mathcal{B}(B\rightarrow K\mu^+\mu^-)}{\mathcal{B}(B\rightarrow Ke^+e^-)}=1+\Big(\frac{2\pi}{\alpha C_9^{SM}}\Big)[(b-1)A_d] R_U \epsilon_l^2(1- 2 s_l^2),
\end{equation}
while the corresponding $\tau/e$ and  $\tau/\mu$ ratios are
\begin{equation}
 R_{K}^{\tau/e} \approx  R_{K}^{\tau/\mu}   \approx \left| 1+\Big(\frac{\pi}{\alpha C_9^{SM}}\Big)[1-(1-a)r_u] R_U \right|^2
 \approx 10^2 \times |1-(1-a)r_u|^2 \times \left(\frac{R_U }{0.1}\right)^2~.
 \label{eq:Rtaumu}
 \end{equation} 
The result in eq.~(\ref{eq:Rtaumu})
holds for any $b\to s\tau\bar\tau$ rate, e.g.~also for $\cB(B\to \tau^+\tau^-)$ and  $\cB(B\to K^* \tau^+\tau^-)$,
normalized to its corresponding SM value. At present this does not represent a significant constraint given that the experimental upper bounds 
on these modes are still 3 orders of magnitude above the SM level~\cite{PDG}, 
but in the future this large enhancement could provide a striking low-energy signature of the model.

 To conclude this section, it is worth comparing the tree-level effects induced by the LQ exchange (model 1) with those 
analyzed in Ref.~\cite{Greljo:2015mma} assuming a vector  uncoloured mediator. 
The structure of the semileptonic operators generated is the same, but the relative weight of charged- and neutral-current terms is different: 
in Ref.~\cite{Greljo:2015mma} the neutral-current operators receive an additional overall suppression factor. 
This is why in Ref.~\cite{Greljo:2015mma} the non-standard effects are much smaller in the case of $b\to s \nu\bar \nu$ and $b\to s   \tau \bar \tau$
transitions. Another difference is the absence, at the tree-level, of LQ contributions to $B$-meson mixing.
However, as we will discuss next, this different is only apparent: quadratically divergent contributions to $\Delta F=2$ amplitudes 
are generated in the LQ case at the one-loop level and, similarly to the case of Ref.~\cite{Greljo:2015mma}, 
 $B$-meson mixing represents a  significant constraint on the model.

\section{Loop effects}

Several processes exist which do not take place in the leptoquark models under consideration at tree level, but appear only at the one loop level. A relevant example, as already noticed, is the $b\rightarrow s \nu\bar\nu$ amplitude in the $SU(2)$-singlet vector leptoquark model 1 of Section 1. 
Since several of these processes can give significant constraints, a one loop calculation is necessary, especially, but not only, when there are quadratically divergent contributions.

For the Lagrangian that describes the free propagation of the leptoquark $U_\mu$ and its interactions with the SM gauge bosons we take
\begin{equation}
\mathcal{L}_U=-\frac{1}{2}U_{\mu\nu}^{\dagger}U^{\mu\nu}+M_U^2 U_{\mu}^{\dagger}U_{\mu}+\mathcal{L}_{an} 
\end{equation}
where 
\begin{equation}
U_{\mu\nu}=D_{\nu}U_{\mu}-D_{\nu}U_{\mu}\quad\quad D_{\mu}\equiv \partial_{\mu}-ig_{s}\frac{\lambda^a}{2}G_{\mu}^a-ig'\frac{2}{3}B_{\mu},
\end{equation}
and
\begin{equation}\label{lqan}
\mathcal{L}_{an}=-ig_{s}k_{s}(U^{\dagger}_{\mu}\frac{\lambda^a}{2}U_{\nu})G^{\mu\nu^{a}}-ig' \frac{2}{3} k_{Y}U^{\dagger}_{\mu}U_{\nu}B^{\mu\nu}
\end{equation}
with obvious meaning of the symbols. $\mathcal{L}_U$ is gauge invariant under the SM group for any value of $k_s$ and $k_Y$. The overall interaction Lagrangian of the leptoquark with the $B_\mu$ field is therefore
\begin{equation}
\mathcal{L}_{UUB}=ig'\frac{2}{3}\Big[(\partial_{\alpha}U_{\beta}^{\dagger}-\partial_{\beta}U_{\alpha}^{\dagger})
B^{\alpha}U^{\beta} - (\partial_{\alpha}U_{\beta}-\partial_{\beta}U_{\alpha})
B^{\alpha}U^{\beta +}
 -k_YU^{\dagger}_{\mu}U_{\nu}
\partial_{\mu}B_{\nu}   + k_Y U_\mu U_\nu^\dagger \partial_\mu B_\nu \Big]
\label{LUUB}
\end{equation}
As a non trivial check of our calculations it will be useful to notice that, for $k_Y=1$ and $2/3 g^\prime = g$, eq.
(\ref{LUUB}) becomes the triple vertex among the $W$ bosons in the SM with the identifications 
\begin{equation}
B_\mu \rightarrow W_{3\mu},\quad\quad U_\mu\rightarrow W_\mu^+,\quad\quad  U_\mu^+\rightarrow W_\mu^-.
\end{equation}
for a fixed color component of the leptoquark field.

\subsection{Quadratically divergent loop effects}

From exchanges of the $U_\mu$ vector quadratically divergent corrections appear: i) in the 2-point function of the $B_\mu$ field; ii) in the 3-point function between $B_\mu$ and the fermion fields; iii) in box-diagram contributions to various 4-fermion interactions.
Below we list the relevant corrections. 

\subsubsection{Two and three-point functions}

In the LQ model we have the following contribution to  the $B_\mu$ propagator
\begin{equation}
  \Pi_{\mu\nu}^{BB}=ig_{\mu\nu}\Big[-k_Y^2\frac{(4/9)g'^2}{64\pi^2}\frac{q^4}{M_U^2} \frac{\Lambda^2}{M_U^2}\Big]
\end{equation}
We do not include a correction to $\Pi_{\mu\nu}^{BB}$ at $q^2=0$, which vanishes by electromagnetic gauge invariance, nor a contribution proportional to $q^2$ since it is reabsorbed in a redefinition of $g^\prime$.

Similarly one has $\Lambda^2$-divergences in the 3-point correlation functions with an external $B_\mu$ field
\begin{equation}\label{2pLQ}
\mathcal{M}_{B\rightarrow  L_3 \bar{L}_3}=-i 3k_Y\frac{(2/3)g'}{64 \pi^2}g_{U}^2\frac{q^2}{M_U^2}\frac{\Lambda^2}{M_U^2} (\bar{u}_{L3}\slashed{\epsilon} P_Lv_{L3}) 
\end{equation}
\begin{equation}\label{3pLQ}
\mathcal{M}_{B\rightarrow  Q_3 \bar{Q}_3} =i k_Y\frac{(2/3)g'}{64 \pi^2}g_{U}^2\frac{q^2}{M_U^2}\frac{\Lambda^2}{M_U^2} (\bar{u}_{Q3}\slashed{\epsilon} P_Lv_{Q3}) 
\end{equation}

As remarked above, this implies the presence in the SM  in the unitary gauge of  similar contributions in the 2 and 3-point functions of the $W_{\mu}^a$ field, at $g^\prime =0$,
\begin{equation}\label{2pSM}
 \Pi_{\mu\nu}^{ab}=ig_{\mu\nu}\delta^{ab} \Big[- \frac{g^2}{64\pi^2} \frac{q^4}{M_W^2} \frac{\Lambda^2}{M_W^2}\Big]
\end{equation}
and
\begin{equation}\label{3pSM}
\mathcal{M}_{W_a \rightarrow  F \bar{F}} = ig \frac{g^2}{64 \pi^2} \frac{q^2}{M_W^2}\frac{\Lambda^2}{M_W^2}(\bar{u}_{F}T^a\slashed{\epsilon}P_Lv_F)
\end{equation}
where $F=Q$ or $L$ and $T^a$ is the  weak isospin. We have explicitly checked that  quadratically divergent contributions to physical amplitudes vanish in this limit of the SM, as they should, with the inclusion of box diagrams as well.

\subsubsection{Box diagrams}
 Some relevant flavour violating processes have  $\Lambda^2$-divergent contributions due to leptoquark box diagrams as well.
The corresponding effective Lagrangian has the form
\begin{equation}
 \mathcal{L}  = \sum_a  F_a  W_a  \frac{g^4_{U}}{64 \pi^2}\frac{\Lambda^2}{M_U^4} O_a
\end{equation}
with the process-dependent effective operators and the corresponding coefficients ($F_a  W_a$)   given in Table \ref{tab:tab1}.

\begin{table}[t]
\begin{center}
\renewcommand\arraystretch{1.6}
\hspace*{-0.05cm}
\begin{tabular}{|c|c|c|c|c|c|}
\hline
Process & Operator & $F_a$ & $W_a$ & Bounds on \\ [-0.3cm]
 &   &  &  & $ F_a \times ( R/0.1)$\\
\hline 
$\tau \rightarrow 3\mu $ &$(\bar{\mu}\gamma^\mu P_L\tau)(\bar{\mu}\gamma_\mu P_L\mu) $ & $3(b-1)^3(c_l\epsilon_l)^3 $&  1 & $ 1.9\times 10^{-2}$ \\
\hline
$\mu \rightarrow 3e$ &$(\bar{e}\gamma^\mu P_L\mu)(\bar{e}\gamma_\mu P_Le) $ & $3(b-1)^4(c_l\epsilon_l)(s_l\epsilon_l)^3 $&  1 &  $5.5 \times 10^{-5}$ \\
\hline
$b \rightarrow s \nu_3\bar{\nu}_3$ &$(\bar{s}\gamma^\mu P_L b)(\bar{\nu}_3\gamma_\mu P_L\nu_3) $ &$[1-(1-a)r_u]$  & $ V_{tb}V^*_{ts}$ & $  [-2.5, 1.4] $\\
\hline
$s \rightarrow d \nu_3\bar{\nu}_3$ &$(\bar{d}\gamma^\mu P_L s)(\bar{\nu}_3\gamma_\mu P_L\nu_3) $ &$[1-(1-a)r_u]^2$ & $V_{ts}V^*_{td}$  & $   [-1.8, 0.6] $ \\
\hline
$b \bar{s}\rightarrow \bar{b}s$ &$(\bar{s}\gamma^\mu P_L b)^2$ &$ [1-(1-a)r_u]^2$ & $(V_{tb}V^*_{ts})^2$  & $3.0 \times 10^{-2} $\\
\hline
\end{tabular}
\caption{Flavour coefficients $F_a \times W_a$  for the  box diagram contributions to the different processes. In the last column are the bounds from current data for $\Lambda = 4\pi M_U/g_U$ (see Section 5.3). The experimental constraints on the various processes are taken from Ref.~\cite{PDG}.}
\label{tab:tab1}
\end{center}
\end{table}

\subsection{Dipole operators}

Unlike the previous cases, the coefficients of the dipole operators are not quadratically divergent. Yet they are logarithmically divergent and yield to 
potentially relevant constraints given the stringent experimental bounds on 
 $\tau \rightarrow \mu \gamma$, $\mu \rightarrow e \gamma$ and $b \rightarrow s \gamma$. The leading LQ 
 contributions to these processes are encoded by 
\be
\mathcal{L}  = \sum_a  F_a   \frac{g^2_U}{32 \pi^2 M_U^2} (1- k_Y) \text{Log}\Big(\frac{\Lambda^2}{M_U^2}\Big) O_a
\ee
where
\ba
&& \mathcal{O}_{\tau \mu  \gamma} = e m_{\tau} (\bar{\mu}_L \sigma^{\alpha\beta}  \tau_R) F_{\alpha\beta}~, \qquad  
F_{\tau\mu}=(b-1)(c_l\epsilon_l)~,  \\
&& \mathcal{O}_{\mu e  \gamma} = e m_{\mu} (\bar{e}_L \sigma^{\alpha\beta}  \mu_R) F_{\alpha\beta}~,  \qquad  
F_{\mu e}=(b-1)^2(c_l\epsilon_l)(s_l\epsilon_l)~, \\
&& \mathcal{O}_{b s  \gamma} = e m_b (\bar{s}_L \sigma^{\alpha\beta}  b_R) F_{\alpha\beta}~, \qquad 
 F_{b s}= \frac{1}{3} V_{tb} V_{ts}^* [1-(1-a)r_u]~. 
\ea

\section{Consistency with data and expected signals}
\subsection{ElectroWeak Precision Tests}

At the one loop level in the leptoquark vector-singlet model there are no corrections to the $S, T, U$ parameters. This is due to the fact that $U_\mu$ only couples to $B_\mu$ and not to the $W_\mu^a$-fields. There are however corrections to the ElectroWeak Precision Tests (EWPT) due to higher dimensional operators, which are at least in principle important due to $\Lambda^2$-divergent effects.

The most effective way to see these effects is by considering the $\epsilon_i$-parameters, as defined in \cite{Altarelli:1990zd}, 
and their expressions in terms of vacuum-polarization, box-diagrams and vertex corrections for $S, T, U= 0$\cite{Barbieri:1991qp}. More specifically, in the limit where one neglects the small $\mathcal{G}_F$-breaking, the first two generations receive corrections only from $\Pi_{\mu\nu}^{BB} = -i g_{\mu\nu}\Pi(q^2)$:
\begin{equation}
\epsilon_1^{(1,2)} = - e_5 
\quad\quad
\epsilon_2^{(1,2)} = - s^2 e_4 - c^2 e_5 
\quad\quad
\epsilon_3^{(1,2)} = c^2 e_4 - c^2 e_5 
\end{equation}
where
\begin{equation}
e_4^{(1,2)} = -\frac{c^2}{2}\frac{d \Pi}{d q^2}(M_Z^2)= - c^2 A,\quad\quad
e_5^{(1,2)}= s^2 \frac{M_Z^2}{2} \frac{d^2\Pi}{d (q^2)^2} = s^2 A,
\end{equation}
and
\begin{equation}
A= k_Y^2\frac{(4/9)g'^2}{64\pi^2}\frac{M_Z^2}{M_U^2} \frac{\Lambda^2}{M_U^2},\quad
 c^2 = 1- \sin^2{\theta_W},\quad\quad \tan{\theta_W}= \frac{g^\prime}{g}
 \end{equation}
To estimate  $A$, we take $\Lambda \approx 4\pi M_U/g_U$, which gives
\begin{equation}
A \approx \frac{k_Y^2}{36 c^2} \left(\frac{g g^\prime}{g_U^2} \right)^2 R_U
\end{equation}
Given the bounds on the deviations from the SM of the $\epsilon_i$ at the $10^{-3}$ level, with $R_U\approx 0.1$ and $g^2_U/(g g') \gtrsim 2\div 3$, this is a mild constraint on $k_Y$. 

Still without any $\mathcal{G}_F$-breaking, a vertex correction intervenes in $Z$-decays to the third generation. Specifically, from eq. (\ref{3pLQ}), the corresponding amplitudes gets corrected as 
\begin{equation}
\delta A_\mu(Z\rightarrow\tau^+ \tau^-) = -i\frac{\delta g}{2} (\bar{u} \gamma_\mu P_L v),\quad\quad
\delta A_\mu(Z\rightarrow b \bar{b}) = i\frac{\delta g}{6} (\bar{u} \gamma_\mu P_L v),
\end{equation}
where
\begin{equation}
\frac{\delta g}{g} = -4 \frac{s^2}{c} k_Y  \frac{g_U^2}{64\pi^2} \frac{M_Z^2}{M_U^2}\frac{\Lambda^2}{M_U^2}
\end{equation}
which can be estimated as
\begin{equation}
\left| \frac{\delta g}{g} \right| \approx s \frac{k_Y}{4 c^2} (\frac{g g^\prime}{g_U^2}) R_U
\end{equation}
Consistency with the few {\it per mille} measurement of $\mathcal{B}(Z\rightarrow \tau^+ \tau^-)$ and $R_U\approx 0.2$ requires 
\begin{equation}
k_Y \lesssim 3\times 10^{-2} \frac{g_U^2}{g g^\prime}.
\end{equation}

\subsection{Box diagrams and dipole operators}

Some of the box diagram contributions shown in Table 1 give extra significant constraints on the parameter space of the  leptoquark vector-singlet model. Such constraints  are given in the last column of the same Table 1 on the modulus of the corresponding flavour coefficients, with the exception of $b\rightarrow s \nu_3 \bar\nu_3$ and $s\rightarrow d \nu_3 \bar\nu_3$,
 where there is an allowed range.  We neglect the contributions, when present,  of  quadratically divergent Penguin-like contributions proportional to $k_Y$.

Although the coefficients of the dipole operators in Section 4.2 are not quadratically divergent, they are significant at least in the leptonic processes. The bound on the flavour coefficient relevant to $\tau\rightarrow \mu \gamma$ (obtained in the limit $k_Y=0$) is
\begin{equation}
(b-1) (c_l\epsilon_l) \log{(\Lambda/M_U)}\lesssim  4\times 10^{-2}  \left(\frac{0.1}{R}\right) 
\end{equation}
whereas the one on $\mu\rightarrow e \gamma$ is
\begin{equation}\label{meg}
(b-1)^2  (s_l c_l)\epsilon_l^2  \log{(\Lambda/M_U)} \lesssim   5\times10^{-5}  \left(\frac{0.1}{R} \right)
\end{equation}

\subsection{Overall constraints}

\begin{figure}[h]
\centering
\includegraphics[scale=0.85]{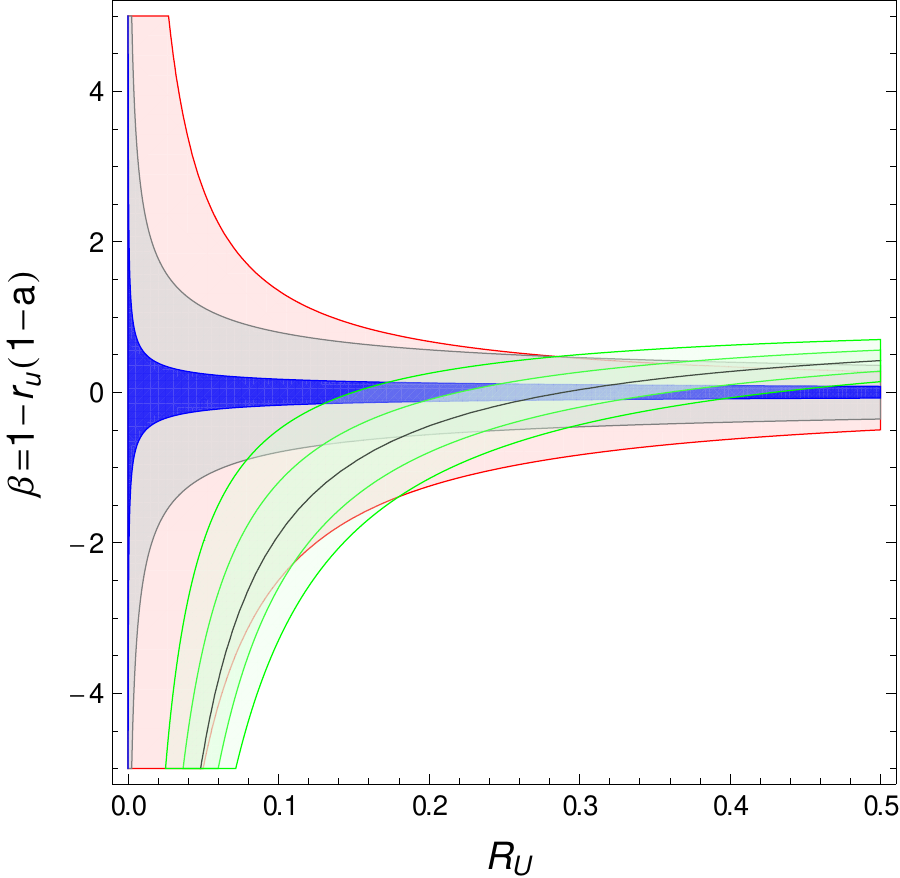}
\caption{Allowed parameter space for the Vector-singlet $U$ from $\Delta B_s=2$ (blue region), $R^{\tau/l}_{D^{(*)}}$ (green), $R_{K^{(*)}\nu}$ (red) and $\mathcal{B}(K^+\rightarrow \pi^+\nu\bar{\nu})$ (gray). $R_U=4g_U^2 M_W^2/g^2M_U^2$ }
\label{fig:qfv}
\end{figure}

As apparent from Sections 3.1 and 5.2 the leptoquark effects in $R_{D^{(*)}}^{\tau/l}$ (at tree level), $R_{K^{(*)}\nu}$, $\mathcal{B}(K^+\rightarrow \pi^+\nu\bar{\nu})$  and $b\bar{s}\rightarrow \bar{b}s$ (al loop level) are predicted in terms of two single effective parameters, $R_U$ and $\beta = 1- (1-a)r_u$, for a cutoff $\Lambda \approx 4\pi M_U/g_U$. The consistency with data of all these effects is shown in Fig. \ref{fig:qfv} for $\Lambda = 4\pi M_U/g_U$. The constraint from $\Delta B_s = 2$ dominates over the others, fixing $\beta$ near zero, within about $0.1\div 0.2$. At the same time, at $1\sigma$ level, $R_U \approx 0.25\div 0.35$. 

\begin{figure}[h]
\centering
\includegraphics[scale=0.85]{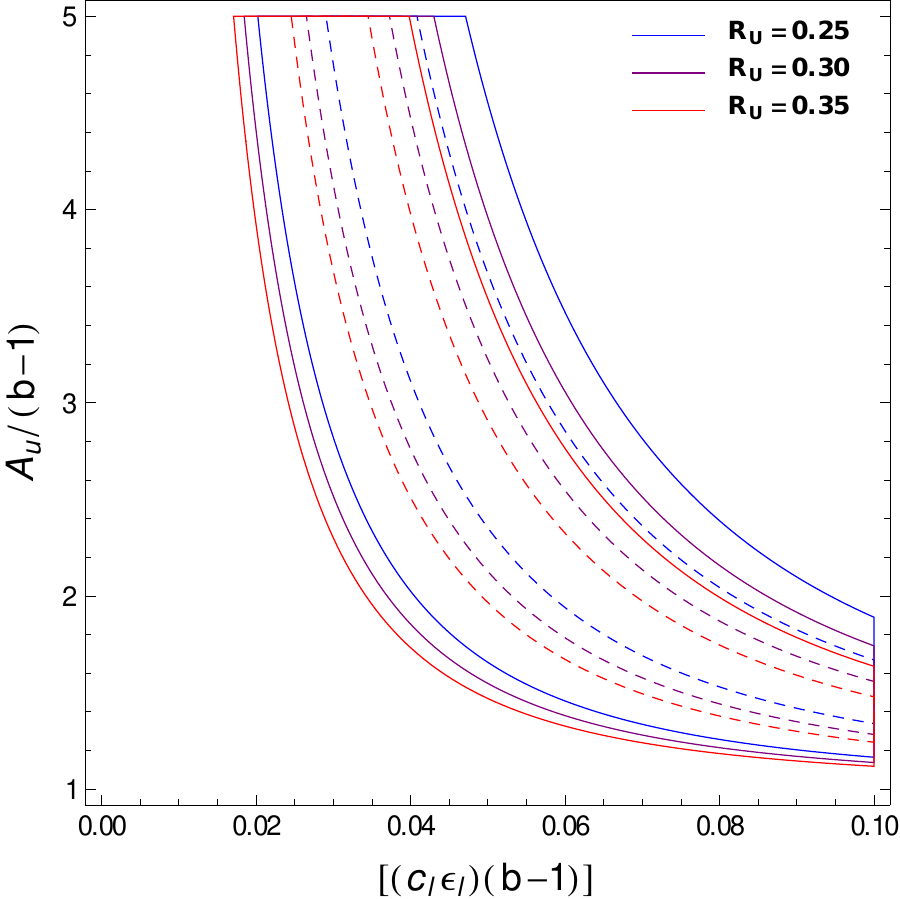}
\caption{Allowed parameter space for the Vector-singlet $U$ from $R_K^{\mu/e}$ for $R_U=0.25, 0.30, 0.35$. Dotted (full) contours delimit the $1\sigma (2\sigma)$ regions.}
\label{fig:neutral}
\end{figure}

\begin{figure}[h!]
\centering
\includegraphics[scale=0.85]{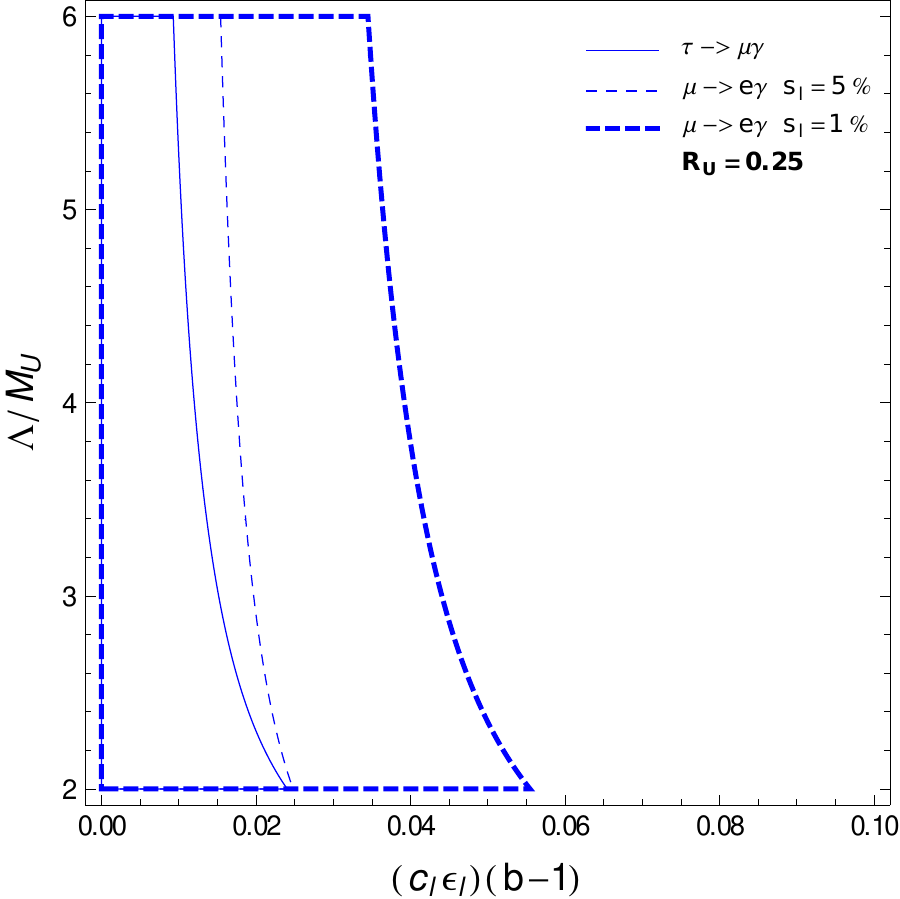}\hspace{7pt}
\includegraphics[scale=0.85]{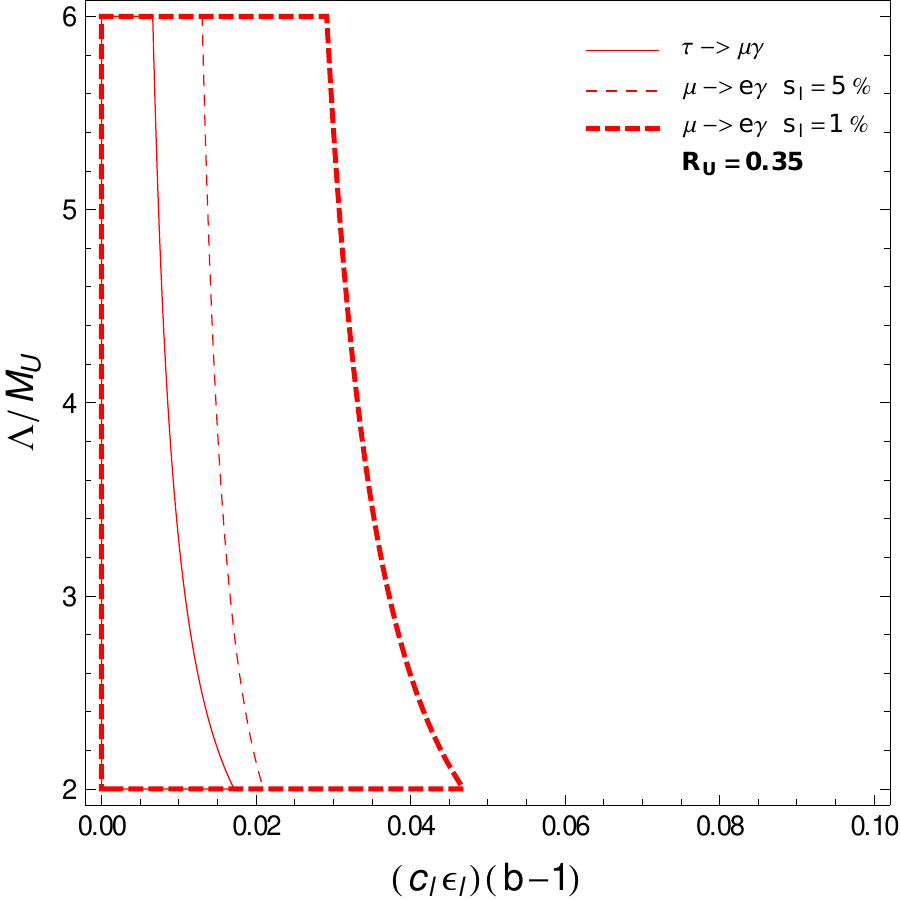}
\caption{Allowed parameter space for the Vector-singlet $U$ from $\tau\rightarrow \mu\gamma$ and $\mu\rightarrow e\gamma$ for $R_U=0.25$ (left) and $R_U=0.35$ (right). }
\label{fig:lfv}
\end{figure}

The leptoquark effects in $R_K^{\mu/e}$ (at tree level) and in the purely leptonic sector (at loop level) depend, for given  $R_U$, on $\epsilon_l$ and $s_l$ plus two combinations of parameters, $(b-1)$ and $A_u =r_u (b-1+a-c)$, generally of order unity. As shown more explicitly in a while, the dipole contribution to $\mu\rightarrow e\gamma$ in eq. (\ref{meg}) requires a small $s_l$. In this case the lepton flavour violating anomaly $R_K^{\mu/e}$ constrains the parameter space as shown in Fig. \ref{fig:neutral}, thus setting a lower bound on the combination $[c_l\epsilon_l (b-1)] \gtrsim 0.02$ for $A_u/(b-1) \lesssim 5$. In turn the dipole contributions to $\tau\rightarrow \mu\gamma$ and to $\mu\rightarrow e\gamma$ (in this case for two values of $s_l$) are shown in Fig.s   \ref{fig:lfv}  against the same combination of parameters and different values of $\Lambda/M_U$. Both Fig. \ref{fig:neutral} and Fig.s   \ref{fig:lfv} are meant to be extended symmetrically for negative values of $[c_l\epsilon_
l (b-1)]$. Note that, for not too small values of $s_l$, it is $\tau\rightarrow \mu\gamma$ that sets the strongest constraint. In fact, for $A_u/(b-1) \lesssim 5$, all anomalies can be  described consistently with all various constraints for values of $\Lambda/M_U$ not larger than about $3\div 4$, suggesting a strong interaction nature of the LQ.

\subsection{Leptoquark pair production at LHC}

The LQ production at the LHC is dominated by the QCD pair production and it depends only 
on $M_U$. A compilation of results relevant to our  model~1 can be found, for instance, in the recent 
CMS analysis~\cite{Khachatryan:2015vaa}: the cross-section varies from about 1 pb for $M_U\approx 0.5$~TeV
to $3\times 10^{-3}$~pb for $M_U\approx 1.0$~TeV\footnote{By taking $k_s=0$ in eq. (\ref{lqan}).}.

In order to determine the LHC sensitivity to various LQ searches we need to evaluate the 
 branching ratios in the different decay channels. At the tree-level one has
\be
\Gamma (U \to q_i \ell_j ) = \frac{1}{24\pi} g_U^2 |F^{U,D}_{ij}|^2 M_U
\ee
and, to a good accuracy, the total decay width is given by the sum of the two leading decays ($U \to t \, \bar\nu_\tau,~b \, \bar \tau$):
$\Gamma_{\rm tot} = g_U^2  M_U/ ( 12\pi)$.
The branching ratios  are then just given by 
\be
\cB(U \to u_i \, \bar\nu_j) = \frac{1}{2} |F^U_{ij}|^2~, \qquad  \cB(U \to d_i \, \ell^+_j) = \frac{1}{2} |F^D_{ij}|^2~.
\ee
For the searches in the second-generation channels performed in Ref.~\cite{Khachatryan:2015vaa}
($U \bar U \to \mu \mu j j$ and $U \bar U \to \mu \nu j j$, where $j$ denote a light-quark jet) we find 
\ba
&&\cB(U \bar U \to \mu\mu j j) = \left( \frac{1}{2}\sum_{i=d,s} |F^D_{i2}|^2 \right)^2  \approx    (8.5 \times 10^{-7} ) \times c_l^4 \epsilon_l^4
 \\
&&\cB(U \bar U \to \mu\nu j j) =  \frac{1}{2} \sum_{i=d,s} |F^D_{i2}|^2
					\times \frac{1}{2} \sum_{j=u,c} \left( |F^U_{j2}|^2 + |F^U_{j1}|^2 \right)
				  \approx (7.7 \times 10^{-7}) \times c_l^2 \epsilon_l^4~, \nonumber 
\ea
from which we deduce that these searches do not put any significant constraint on the model.

On the other hand, a relevant constraint is obtained by the dedicated search for the 
$U \bar U \to t \bar t \nu_\tau \bar\nu_\tau $ decay chain performed by ATLAS~\cite{Aad:2015caa}.
In our model  $\cB(U \bar U \to t \bar t \nu_\tau \bar\nu_\tau)=0.25$ that implies the limit 
\begin{equation}
M_U > 770~\GeV~.
\end{equation}
By a naive scaling of the statistics  and the cross-section, we estimate that this limit could improve 
up to 1.3~TeV, in absence of a signal, with 300~fb$^{-1}$ at 13~TeV.

Taking into account the constraints on $R_U$ in Fig. 2, the bound on $M_U$ can be turned into 
a bound on $g_U$. For $R_U> 0.2$ we get $g_U > 1.4$, that would raise to  $g_U > 2.4$ in absence
of a direct signal  with 300~fb$^{-1}$ at 13~TeV.

\section[A \textit{naive} composite leptoquark picture]{A \textbf{\textit{naive}} composite leptoquark picture}

Needless to say the phenomenological model described so far cries out for a UV completion. Here we describe an attempt in the direction of composite Higgs models, that will at least serve to illustrate the difficulties to comply with the various constraints. It does not take much to anticipate that the main such constraint comes from the value of $R_U$ as implied by Fig. 2.

Let us consider a strongly interaction sector with a global symmetry $SU(4)\times SO(5)$ spontaneously broken down to the Pati-Salam group $SU(4)\times SU(2)_L\times SU(2)_R$, so as to generate a composite pseudo-Goldstone Higgs boson. As usual the SM group is gauged inside the residual global group. Within $SU(4)$ this structure leads to composite quasi-degenerate vectors: the composite gluons, a vector singlet carrying $B-L$ and, most importantly,  the $U_\mu$ leptoquark vector. 

The strong sector will also contain composite vector-like fermions in multiplets of the Pati Salam group occurring in three flavour species. The important thing is that these composite $\Psi$ fermions have components that match the quantum numbers of the SM fermions with respect to the SM gauge group. The requisite to make contact with the phenomenological model described in the previous Sections is that the mass mixing terms between the elementary and the composite fermions respect, up to small breaking terms, the symmetry in flavor space $U(2)_\Psi \times \mathcal{G}_F$\footnote{If $\Psi$ is reducible $U(2)_\Psi$ may actually be a product of $U(2)$ factors}. 
With $\Psi = (4, 2,1) \oplus (\bar{4}, 1, 2)$ (plus complex conjugate states)
it is easy to convince oneself that  this structure shall lead, in the unbroken $U(2)_\Psi \times \mathcal{G}_F$ limit, to eq. (\ref{L1}) with
\begin{equation}
g_U= g^* \sin{\theta_{Q3}} \sin{\theta_{L3}}
\end{equation}
and $F_{ij} = \delta_{i3} \delta_{j3}$ as the only interaction of $U_\mu$ with the standard fermions. Here $g^*$ is the  coupling of the LQ to the composite fermions and $ \theta_{Q3},  \theta_{L3}$ are the mixing angles of $Q_3, L_3$ with the same composite fermions in $\Psi$. Other choices of $\Psi$, phenomenologically motivated, shall lead to a relation between $g_U$ and $g^*$ involving more parameters but always respecting $g_U\leq g^*$.

The "standard" interpretation of  the composite Higgs models also gives a similar mass $M \approx g^* f$ for the composite vectors in the adjoint of $SU(4)$, among which is $U_\mu$, and $f$ is the breaking scale of $SO(5)$ down to $SO(4)$. As a consequence  the phenomenological parameter $R_U$ becomes
\begin{equation}
R_U \approx  \left(\frac{V}{f} \right)^2  \sin^2{\theta_{Q3}} \sin^2{\theta_{L3}},\quad\quad V = 245~GeV,
\end{equation}
independent from $g^*$. 

\section{Summary and conclusions}

Being the heaviest particle in the SM, the top quark plays a special role in many processes and/or mechanisms of the greatest relevance in particle physics. It is not surprising therefore that the top quark is thought to be equally important in several BSM speculative considerations. Flavour physics is no exception to this rule. In the SM top exchanges dominate many of the observed Flavour Changing Neutral Current effects. In BSM, taking Minimal Flavour Violation as a relevant example, it is the relatively large top Yukawa coupling that controls many of the new observable phenomena that might occur.

This is the basis to consider a $U(2)^3$ approximate symmetry as a ruling symmetry of the flavor quark sector of any putative extension of the SM. More precisely $U(2)_Q\times U(2)_u\times U(2)_d\times U(1)_{d3}$, a residual  symmetry of the SM with all the Yukawa coupling switched off but the top one, can allow, suitably broken, for mild deviations from the SM itself.  Decays of the $B$ mesons, in particular through the $b_L$-component, which is the only singlet under $U(2)^3\times U(1)_{d3}$ other than $t_{L,R}$, are obvious candidates where such deviations might occur and be observable.

It is therefore natural to ask if and how the recently emerged anomalies in semileptonic $B$-decays can be accommodated in this context. As we have shown this is possible by:
\begin{itemize}
\item[i.)] invoking the exchange of a vector-singlet LQ with a relatively large value of the parameter $R_U = 4M_W^2 g_U^2/g^2 M_U^2\approx 0.2$, i.e. a
ratio well below 1 TeV between its mass $M_U$ and its dominant coupling $g_U$ to the third generation of left-handed quarks and leptons;
\item[ii.)]  extending the $U(2)^3\times U(1)_{d3}$ symmetry of the quark sector, with its breaking, to the lepton sector as well, via a $U(2)_L\times U(2)_e\times U(1)_{e3}$ flavour symmetry.
\end{itemize}
The observed anomalies arise as relatively small effects of the breaking of the overall $U(2)^5$   symmetry, thus allowing non vanishing couplings of the LQ to the lighter generations as well. 

The deviation from the SM in the charged current observable $R^{\tau/l}_{D^{(*)}}$ is due to a tree level exchange of the LQ controlled, other than by $R_U$, by a single combination of dimensionless  parameters $\beta$, as shown in Fig. \ref{fig:qfv}. These same effective parameter controls the quadratically divergent one loop contributions to $B\rightarrow K \nu_3\bar\nu_3$, $K\rightarrow \pi \nu_3\bar\nu_3$ and $ b \bar{s}\rightarrow \bar{b} s$, as well as the tree level effect in $b\rightarrow s \tau \bar{\tau}$.
All of these processes can corroborate or exclude the LQ model by future measurements. 

The lepton flavour violation  emerging in the neutral current observable $R_K^{\mu/e}$
can also arise from a tree level exchange of the leptoquark, although suppressed by the intervention in the final state of muons or electrons. In this case the relevant parameters control as well the log-divergent one loop dipole moment contributions to $\tau\rightarrow \mu \gamma$  and $\mu\rightarrow e \gamma$. The current limits on both these processes are close to saturate the observed deviation  of  $R_K^{\mu/e}$ from unity, at least in a natural range of the relevant parameters, as illustrated in Fig. \ref{fig:neutral} and Fig.s   \ref{fig:lfv}.

To discuss the possible  manifestation of the vector LQ in direct production strongly depends on being able to go beyond the  low energy phenomenological picture that we have used in this paper. This is because  the constraint on $R_U\approx 0.2$ only fixes the ratio $M_U/g_U$ between the mass and the coupling of the LQ. The current LHC searches bound $M_U$ to be bigger than about 700 GeV, independently from the value of $g_U$, which is therefore constrained to be bigger than about 1.4. Larger values of $g_U$, and therefore of $M_U$ as well, are indirectly hinted  by the constraints from $\tau\rightarrow \mu \gamma$  and $\mu\rightarrow e \gamma$.
All this points towards the need of a UV completion of the phenomenological model used so far, perhaps along the lines outlined in Section 6. 
It will be interesting to see if and how such UV completion can be accomplished consistently with the numerous constraints.

\subsubsection*{Acknowledgments}

We thank Andrea Tesi and Andrea Wulzer for useful discussions.
This research was supported in part by the Swiss National Science Foundation (SNF) under contract 200021-159720.

\end{document}